\documentclass[prd,singlecolumn,superscriptaddress,showpacs,nofootinbib,
preprintnumbers]{revtex4}

\usepackage{amsmath}
\usepackage{amsfonts}
\usepackage{graphicx}
\usepackage{dcolumn}

\def\be{\begin{equation}}
\def\ee{\end{equation}}
\def\ba{\begin{eqnarray}}
\def\ea{\end{eqnarray}}
\def\bs{\begin{subequations}}
\def\es{\end{subequations}}

\usepackage{color}


\begin{document}

\title{Constraining $f(R)$ gravity models with disappearing cosmological constant}
\author{I. Thongkool}
\affiliation{Centre of Theoretical Physics, Jamia Millia Islamia,
New Delhi-110025, India}
\author{M. Sami}
\affiliation{Centre of Theoretical Physics, Jamia Millia Islamia,
New Delhi-110025, India}
\author{R. Gannouji}
\affiliation{IUCAA, Post Bag 4, Ganeshkhind, Pune 411 007, India}
\author{S. Jhingan}
\affiliation{Centre of Theoretical Physics, Jamia Millia Islamia,
New Delhi-110025, India}
\begin{abstract}
The $f(R)$ gravity models proposed by Hu-Sawicki and Starobinsky are
generic for local gravity constraints to be evaded. The large
deviations from these models either result into violation of local
gravity constraints or the modifications are not distinguishable
from cosmological constant. The curvature singularity in these
models is generic but can be avoided provided that proper fine
tuning is imposed on the evolution of scalaron in the high curvature
regime. In principle, the problem can be circumvented by
incorporating quadratic curvature correction in the Lagrangian
though it might be quite challenging to probe the relevant region
numerically.
\end{abstract}

\maketitle
\section{Introduction}
The growing faith in the late time cosmic acceleration is directly
supported by observations of high red-shift supernovae  and
indirectly by observations on microwave background, large scale
structure and weak lensing. What causes the repulsive  effect, in
the cosmic expansion, is one of mysteries of modern cosmology at
present. Theoretically, the phenomenon can be accounted for either
by supplementing the energy momentum tensor by an exotic matter
component with large negative pressure ({\it dark
energy})\cite{review1,vpaddy,review2,review3,review3C,review3d,review4}
or by modifying gravity itself. Cosmological constant, the simplest
candidate of dark energy, is plagued with fine tuning problem of an
unacceptable level\cite{rev0}.
 Scalar field could provide an interesting alternative to
cosmological constant\cite{review2}. They can mimic cosmological
constant like behavior at late times and can give rise to a viable
cosmological dynamics at early epochs. Scalar field models with
generic features are capable of alleviating the fine tuning and
coincidence problems\cite{review2}. As for the observations, at
present, they are absolutely consistent with $\Lambda$ but at the
same time, a large number of scalar field models are also permitted.
Future data should allow to narrow down the class of permissible
models of dark energy.

As an alternative to dark energy, the large scale modifications of
gravity could account for the current acceleration of universe. We
know that gravity is modified at short distance and there is no
guarantee that it would not suffer any correction at large scales
where it is never verified directly. Large scale modifications might
arise from extra dimensional effects or can be inspired by
fundamental theories of high energy physics. On purely
phenomenological grounds, one could seek a modification of Einstein
gravity by replacing the Ricci scalar in Einstein-Hilbert action by
$f(R)$\cite{FRB}. However, any large scale modification of gravity
should reconcile with Local Gravity Constraints and should have
potential of being distinguished from cosmological constant. Since
the general theory of relativity is in excellent agreement with
local gravity phenomenon, it is quite challenging to construct a
viable model of $f(R)$ gravity along the said lines. Stability
requires that the first and the second derivatives of $f(R)$ with
respect to the Ricci scalar $R$ should be positive definite. Most of
the corresponding modifications of the Einstein-Hilbert action are
either cosmologically un-viable or can not be distinguished for the
cosmological constant.

 The class of models proposed by
Hu-Sawicki and Starobinsky (HSS) is of great interest. These models
can evade local gravity constraints and have potential capability of
being distinguished from the cosmological constant\cite{HS,star}(see
also Ref.\cite{Apl1}). However, they are quite delicate $-$ the
minimum of the scalaron (scalar degree of freedom present in $f(R)$
gravity) potential which corresponds to dark energy in these models
is very near to field configuration corresponding to infinitely
large value of $R$ for solar physics constraints to evaded. Thus it
is quite likely that the scalar field, which controls the space-time
curvature, hits singularity while evolving near the de-Sitter
minimum\cite{Apl2,frolov,TH,maeda1,maeda2,NOOD}. The problem becomes
acute in high curvature regime but can be circumvented by carefully
tuning the parameters of the model\cite{Lango,whu}.

The HSS models are characterized by a finite potential barrier
between the minimum of the scalaron potential and the curvature
singularity and hence are vulnerable to singularity. Recently, the
HSS models were modified such that the said potential barrier is
infinite and the curvature singularity is hidden behind the infinite
potential barrier\cite{waga}.

In this paper, we examine the deformations of HSS models and
demonstrate that these models are generic to local gravity
constraints. We also argue that the viable resolution of curvature
singularity can be provided by adding higher curvature terms to the
originally proposed form of $f(R)$ in Refs.\cite{HS,star}.

\section{Large curvature singularity versus the local gravity constraints}
The action of  $f(R)$ gravity in Jordan frame in the presence of
matter described by the matter Lagrangian ${\cal L}_m$ is given by
\cite{review1},
\begin{equation}
S = \int{\rm d}^{4}x\sqrt{-g}\left[\frac{f(R)}{2} + \mathcal{L}_m \right], \label{actionJ}
\end{equation}
where the matter Lagrangian depends on the metric $g_{\mu \nu}$ and
the matter fields. In what follows, it would be convenient to write
$f(R)$ in the following form,
\begin{equation}
\label{eq:psi}
 f(R)=R+\Delta, \qquad \psi = \frac{\partial f}{\partial R} = 1 + \Delta_{,R},
\end{equation}
where $\Delta$ describes the correction to Einstein-Hilber action
and $\Delta_{,R} $ denotes its derivative with respect to the Ricci
scalar $R$. The $f(R)$ theory apart from the spin-2 object
necessarily contains a scalar degree of freedom which becomes clear
either by taking the trace of the modified Einstein equations
obtained from (\ref{actionJ}) or by passing to the Einstein frame.
Indeed one can always make a conformal transformation which
converts the original action (\ref{actionJ}) into Einstein-Hilbert
action along with a canonical scalar field $\phi$ which directly
couples to matter. The solar system and equivalence principle bounds
give a strong constraint on the magnitude of the scalar field $\phi$
in the Einstein frame. The potential of field $\phi$ is uniquely
constructed from the Ricci scalar $R$.

We now transform the metric using the conformal transformation,
\begin{equation}
 \tilde{g}_{\mu\nu}=\psi~g_{\mu\nu},\qquad \phi=\sqrt{\frac{3}{2}}\ln \psi.
 \label{phi}
\end{equation}
The action in the Einstein frame is given by
\begin{equation}
 S = \int{\rm d}^4 x \sqrt{-\tilde{g}}\left [
\frac{\tilde{R}}{2}-(\tilde{\nabla}\phi)^2-V(\phi)+{\cal L}_m(\tilde{g}_{\mu
\nu}e^{2g_c\phi}) \right ],
\end{equation}
where the coupling $g_c$ and V are given by,
\begin{equation}
 g_c = -\frac{1}{\sqrt{6}}, \qquad V = \frac{Rf_{,R}-f}{2f_{,R}^2}.
\end{equation}
As shown in Ref.\cite{khoury},  the thin shell parameter is given by
\begin{equation}
 \frac{\Delta \tilde{r}_c}{\tilde{r}_c} = \frac{\phi_B - \phi_A}{6 g_c \Phi_c},
 \label{thinshell}
\end{equation}
where $\phi_A,\phi_B$ are corresponding to the minimum of the effective
potential
\begin{equation}
 V_{eff}(\phi)=V(\phi)+e^{g_c \phi}\rho^*,
 \label{Veff}
\end{equation}
inside and outside the spherical body respectively and $\Phi_c$ is
the gravitational potential of the test body (Sun/Earth).
Let us consider the  variants of HSS models\cite{waga},
\begin{equation}
\Delta = \alpha \beta R_c\left( \left [ 1 + \left ( \frac{R}{R_c}
\right )^n \right ]^{-1/\beta }-1\right),~~R_c>0
\label{ES}
\end{equation}

The conditions for the cosmological viability of $f(R)$ models can be understood by considering two quantities \cite{luca}:

\be m=\frac{R f_{,RR}}{f_{,R}},\quad r=-\frac{Rf_{,R}}{f} \ee

The presence of a viable saddle matter era demands that

\be
m(r\approx -1) \approx 0, \quad m'(r\approx -1)>-1
\label{conditionsM}
\ee

 The conditions (\ref{conditionsM}) are
satisfied for the model (\ref{ES}) provided that

\be
n>0, \quad \text{and}~~(\beta>0,~~\text{or}~~\beta<-n)
\ee

In what follows we shall consider the case of $(n,\beta)>0$. In fact
we know \cite{capozziello} that for $n>0$ and $\beta<0$ the model is
not distinguishable  from the $\Lambda$CDM.


Let us emphasize that HSS models in Starobinsky parametrization
corresponding to $n=2$ and $\beta\leq  1$ has a moderate dependence
on $R$ allowing the local gravity constraints to be evaded.

Let us now analyse extended HSS models\cite{waga} described by
(\ref{ES}). In this case, in the high curvature regime $R \gg R_c$,
we obtain,
\begin{equation}
 \Delta_{,R} \approx - \alpha n \left ( \frac{R}{R_c} \right )^{-\frac{n}{\beta}-1},
\end{equation}
which shows that $\Delta_{,R} \ll 1$ in the case under consideration
($n,\beta>0$) for moderate values of $\alpha$. Using expression for
$\phi$ given by Eq.(\ref{phi}) and the fact that $\Delta_{,R}\ll 1$, we
find,
\begin{equation}
 \phi = \sqrt{\frac{3}{2}}\ln (1+\Delta_{,R}) \approx \frac{\sqrt{6}}{2} \Delta_{,R},
\end{equation}
We next estimate $R$ corresponding to minimum of the effective
potential,
\begin{equation}
 \frac{dV_{eff}}{d\phi}=-g_c \left [ \frac{R(1-\Delta_{,R})+2\Delta}{(1+\Delta_{,R})^2}\right ]
+g_ce^{g_c\phi}\rho^* =0
\end{equation}
which simplifies in case of the generic approximation, $\Delta_R \ll
1, \Delta \ll R$ and gives rise to following expression for
$\phi_{min}$
\begin{equation}
 \phi_{min} \approx \frac{\sqrt{6}}{2} \Delta_{,R}|_{R=\rho^*}\approx
-\frac{\sqrt{6}}{2}\alpha n \left ( \frac{R_c}{\rho^*} \right
)^{\frac{n}{\beta}+1}. \label{phimin}
\end{equation}
Hereafter, we shall use the notation $\rho$ for matter density
instead of $\rho^*$ in Einstein frame. From the fact that $\rho_A$,
the energy density inside the test bodies (Sun/Earth) is of the
order of 1 g/cm$^{3}$ which is much larger than the density outside
($\rho_B \sim 10^{-24}$ g/cm$^{3}$ of the baryonic/dark matter
density in our galaxy), it follows from Eq.(\ref{phimin})that
$|\phi_A| \ll |\phi_B|$,
\begin{equation}
|\frac{\phi_A}{\phi_B}|\simeq
\left(\frac{\rho_B}{\rho_A}\right)^{\frac{n}{\beta}+1}\ll 1
\end{equation}
which allows us to write the thin shell condition in the convenient
form
\begin{eqnarray}
\label{eq:phiconstraint}
 |\phi_B|
&\simeq& \sqrt{6}\Phi_c\frac{\Delta {\tilde r}_c}{{\tilde r}_c}, \\
&\lesssim& \left \{ \begin{array}{rl}
 5.97\times 10^{-11} &\textrm{(Solar system test)}, \\
 3.43\times 10^{-15} &\textrm{(Equivalence Principle (EP) test)}.
\end{array} \right .
\end{eqnarray}
We have used $\frac{\Delta {\tilde r}_c}{{\tilde r}_c} < 1.15 \times
10^{-5}$, $\Phi_c \simeq 2.12\times 10^{-6}$ for the Sun and
$\frac{\Delta {\tilde r}_c}{{\tilde r}_c} < 2 \times 10^{-6}$,
$\Phi_c \simeq 7\times 10^{-10}$ to respect the equivalence
principle constraint. In what follows, we shall investigate the
modified HSS models (\ref{ES}) for different values of model
parameters. Let us first consider the case of $\beta \to \infty$ and
$n = 1$ \cite{waga},
\begin{equation}
 f(R) = R - \alpha R_c \ln \left ( 1 + \frac{R}{R_c} \right )
\quad \Longrightarrow \quad \phi_B \approx
\frac{\sqrt{6}}{2}\Delta_{,R}\big |_{R = \rho_B} =
-\frac{\sqrt{6}}{2}\frac{\alpha R_c}{R_c+\rho_B},
\end{equation}
The de-Sitter minimum  in free space is given by
\begin{equation}
 \frac{dV}{d\psi}=\frac{1}{2\psi^3}\left ( 2f-R\psi \right )
= \frac{1}{2(1+\Delta_{,R})^3}\left ( R+2\Delta-R\Delta_{,R} \right ) = 0.
\end{equation}
which gives rise to the following relation for $\alpha$
\begin{equation}
\alpha = \frac{x_1(1+x_1)}{-x_1+2(1+x_1)\ln (1+x_1)},~~x_1\equiv
\frac{R_1}{R_c} \label{alpha}
\end{equation}
Relation (\ref{alpha}) implies that $\alpha$ is always positive
definite for any $x_1$ and that $\alpha\to 1$ as $x_1\to 0$. For
moderate values of $\alpha$, the de-Sitter minimum corresponds to
$R_1 \sim \rho_c$ ($\rho_c \simeq 10^{-29}$ g/cm$^{-3}$). For
instance, in case of $\alpha=2$, we find that, $R_1\simeq 6 R_c$
which gives the estimate for $|\phi_B|$ as $|\phi_B|\gtrsim
10^{-6}$. This is clearly ruled out by the thin shell condition
(\ref{thinshell}).

We next investigate the model (\ref{ES}) for arbitrary values of
parameters.

\subsection{Constraint for the general $\beta,n,\alpha$}

Let us define the dimensionless variable $x$ as $x \equiv R/R_c$ and
write the expression of interest in terms of $x$,
\begin{eqnarray}
 \Delta &=& -\alpha \beta R_c \left \lbrace 1 - (1+x^n)^{-1/\beta} \right \rbrace,\\
 \Delta_{,R} &=& -n\alpha x^{n-1}(1+x^n)^{-1/\beta -1},\\
V &=& \frac{R\Delta_{,R}-\Delta}{2(1+\Delta_{,R})^2}, \\
  &=& -\frac{\alpha R_c}{2}\frac{(1+x^n)^{-1/\beta-1}
\left \lbrace nx^n-(1+x^n)\left [ -1 + (1+x^n)^{1/\beta} \right
]\beta \right \rbrace} {\left [ -1+n\alpha
x^{n-1}(1+x^n)^{-1/\beta-1} \right ]^2}
\end{eqnarray}
The de-Sitter minimum in free space in this case corresponds to
\begin{eqnarray}
 \alpha &=&
\frac{x_1(1+x_1^n)^{1+1/\beta}}{-nx_1^n+2(1+x_1^n)\left [ -1 +
(1+x_1^n)^{1/\beta} \right ]\beta},
\end{eqnarray}

For $\beta \to \infty$, these equations reduce to
\begin{eqnarray}
 \Delta &=& -\alpha R_c \ln (1+x^n) ,\\
 \Delta_{,R} &=& -\frac{n\alpha x^{n-1}}{x^n+1},\\
V &=&  -\frac{\alpha R_c}{2}\frac{x^2(1+x^n)\left [ nx^n-(1+x^n)\ln
(1+x^n)\right ]}
{(x+x^{n+1}-\alpha n x^n)^2} \\
\alpha &=& \frac{x_1(1+x_1^n)}{-nx_1^n+2(1+x_1^n)\ln (1+x_1^n)}
\end{eqnarray}

\subsection{case: $\beta \to \infty$ and $n\geq 2$}
Our numerical analysis shows that $\alpha$ is positive definite for
all values of $x_1$ provided that $n<2$. However, for larger values
of $n$  there exist values of $x_1$ for which $\alpha$ is positive.
In case of $n \gtrsim 10$ corresponding  to $x_1
> \sqrt{e} = 1.649$, the parameter, $\alpha$ is
always positive as shown in Fig.\ref{binfinity}. In this case
 $x_1^n \gg 1 $ and we obtain
\begin{equation}
 \alpha = \frac{x_1(1+x_1^n)}{-nx_1^n+2(1+x_1^n)\ln (1+x_1^n)}
\simeq \frac{x_1^{n+1}}{x_1^n(2\ln(x_1^n)-n)} \simeq
\frac{x_1}{n(2\ln x_1 -1)}, \label{alpha2}
\end{equation}
which we shall use to confront the model with solar tests
\begin{equation}
|\phi_B|\approx -\frac{\sqrt{6}}{2}\Delta_{,R}{\big |}_{R=\rho_{_B}}
=\frac{\sqrt{6}}{2}\frac{n\alpha\left (\frac{\rho_B}{R_c}\right
)^{n-1}} {\left (\frac{\rho_B}{R_c} \right)^n+1} \sim
\frac{\sqrt{6}}{2}n\alpha\left (\frac{\rho_B}{R_c} \right )^{-1}
\sim \frac{\sqrt{6}}{2} n\alpha \left (x_1 \frac{\rho_B}{R_1}
\right)^{-1} \sim \frac{\sqrt{6}}{2}\frac{n\alpha}{x_1}\times
10^{-5}, \label{phiB1}
\end{equation}
Substituting, $\alpha$ from Eq.(\ref{alpha2}) in (\ref{phiB1}), we
have
\begin{equation}
|\phi_B| \approx \frac{1}{2\ln x_1 - 1}\times 10^{-5}
\end{equation}
To satisfy, $|\phi_B| < \mathcal{O}(10^{-10})$ (EP constraint), we
need $\ln x_1$ to be very large number which implies that the model
can not be distinguished from $\Lambda CDM$.
\subsection{case: $\beta \to \infty,n \to 0 $}
\begin{equation}
 \alpha = \frac{x_1(1+x_1^n)}{-nx_1^n+2(1+x_1^n)\ln (1+x_1^n)}
\sim\frac{x_1(1+1)}{(1+1)\ln2} \sim \frac{x_1}{\ln2}
\end{equation}
and
\begin{equation}
|\phi_B| \approx
\frac{\sqrt{6}}{2}n\alpha\left ( \frac{\rho_B}{R_c} \right )^{-1}
\sim \frac{\sqrt{6}}{2}\frac{n\alpha}{x_1}\times 10^{-5}
\sim \frac{\sqrt{6}}{2}\frac{n}{\ln 2}\times 10^{-5}
\end{equation}
then $n< \mathcal{O}(10^{-10}$), as implied by the EP constraint,
which makes the model  indistinguishable from cosmological constant.

\subsection{case: $\beta \to \infty,n<2$}
\begin{figure}
\includegraphics[width=11cm,angle=-90]{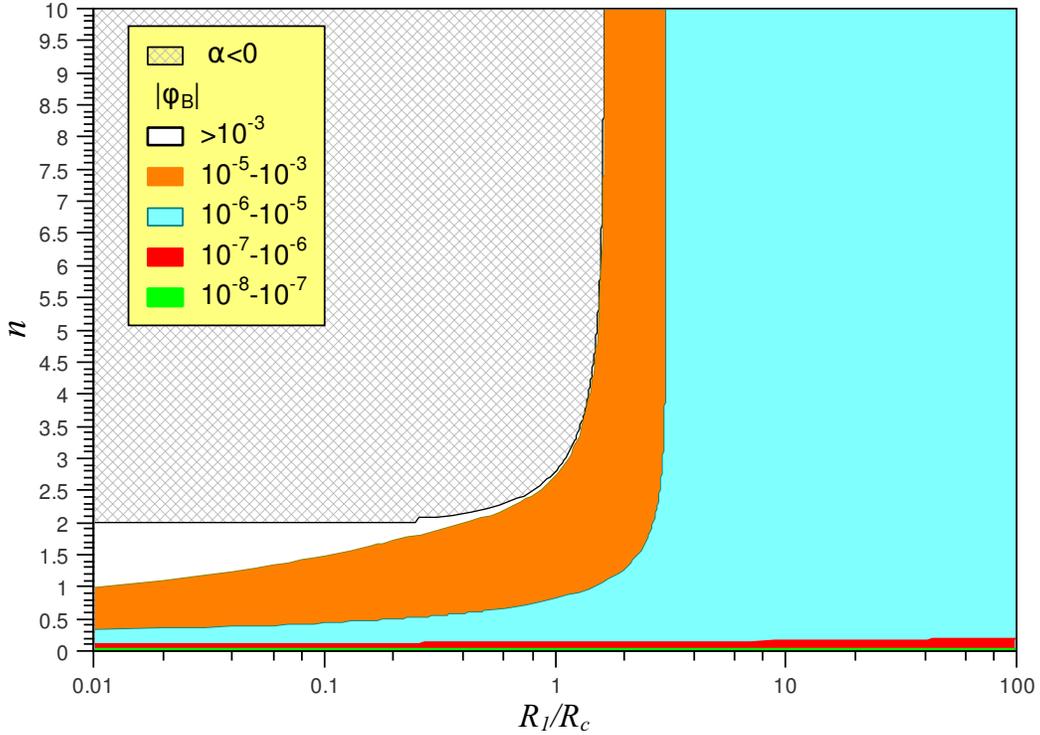}
\caption{Plot of the parameter space for $R_1/R_c$ varying from 0.01
to 100 and $n$ ranging from 0.005 to 10. The plot shows that there
is no region in this parameter space for the case $\beta \to \infty$
for the local gravity constraints to be satisfied, $|\phi_B|
\lesssim 10^{-10}$.} \label{binfinity}
\end{figure}

As shown above, for arbitrary value of $n$, the parameter $\alpha$
is positive definite provided that $n<2$ for all values of $x_1$.
This can easily be demonstrated analytically in the limit of $x_1
\to 0$,
\begin{equation}
 \alpha = \frac{x_1(1+x_1^n)}{-nx_1^n+2(1+x_1^n)\ln (1+x_1^n)}
\sim\frac{x_1}{-nx_1^n + 2(1)(x_1^n)} \sim \frac{x_1^{1-n}}{2-n}
\end{equation}
In the region of positive $\alpha$ and $0.2<n<2$, our numerical
estimates show (see Fig.\ref{binfinity}) that $|\phi_B|\gtrsim
10^{-6}$. As mentioned before, the model may be compatible with
solar test for $n\lesssim 10^{-10}$ but  reduces to $\Lambda CDM$.
 Let us note that the class of
models\cite{waga}
\begin{equation}
\Delta=-\alpha R_c\left(1+\frac{R}{R_c}\right)^n
\end{equation}
is practically not distinguishable from cosmological constant as the
local gravity constraints impose severe restriction on $n$, namely,
$n<10^{-10}$.

\begin{figure}
\includegraphics[width=11cm,angle=-90]{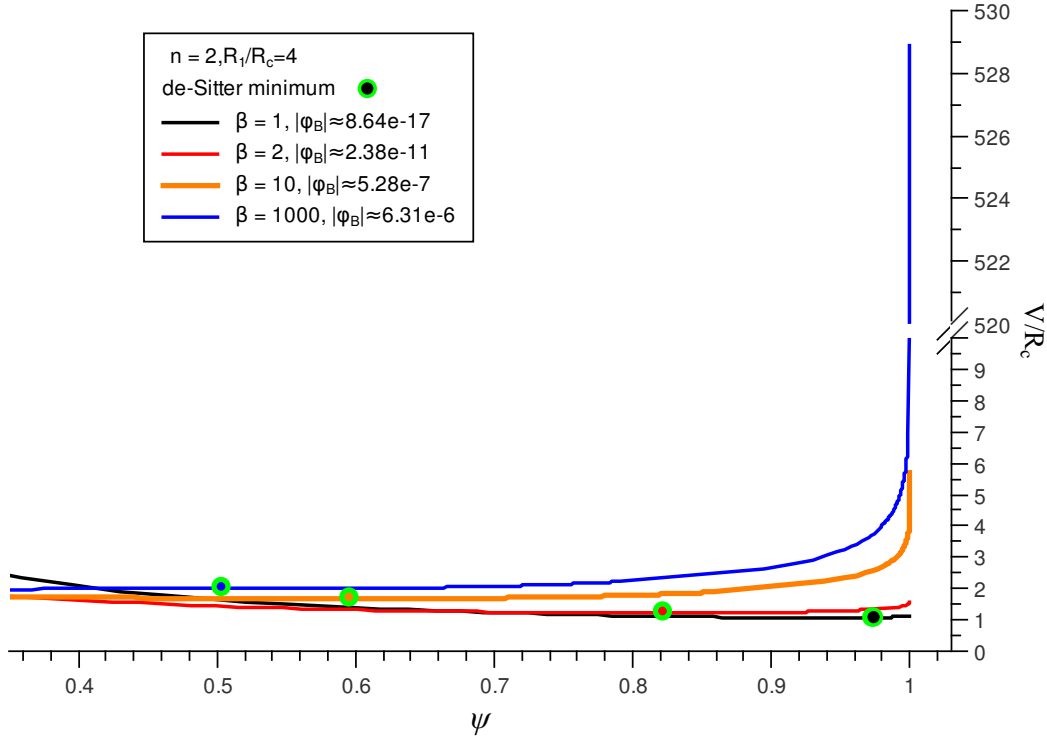}
\caption{Plot of the potential $V$ versus $\psi$ for different values
of parameters. The plot shows that the potential barrier becomes
large for the large values of $\beta$. The de-Sitter minimum is also
seen to shifts towards the singularity as $\beta$ decreases.}
\label{pott}
\end{figure}
\begin{figure}
\includegraphics[width=11cm,angle=-90]{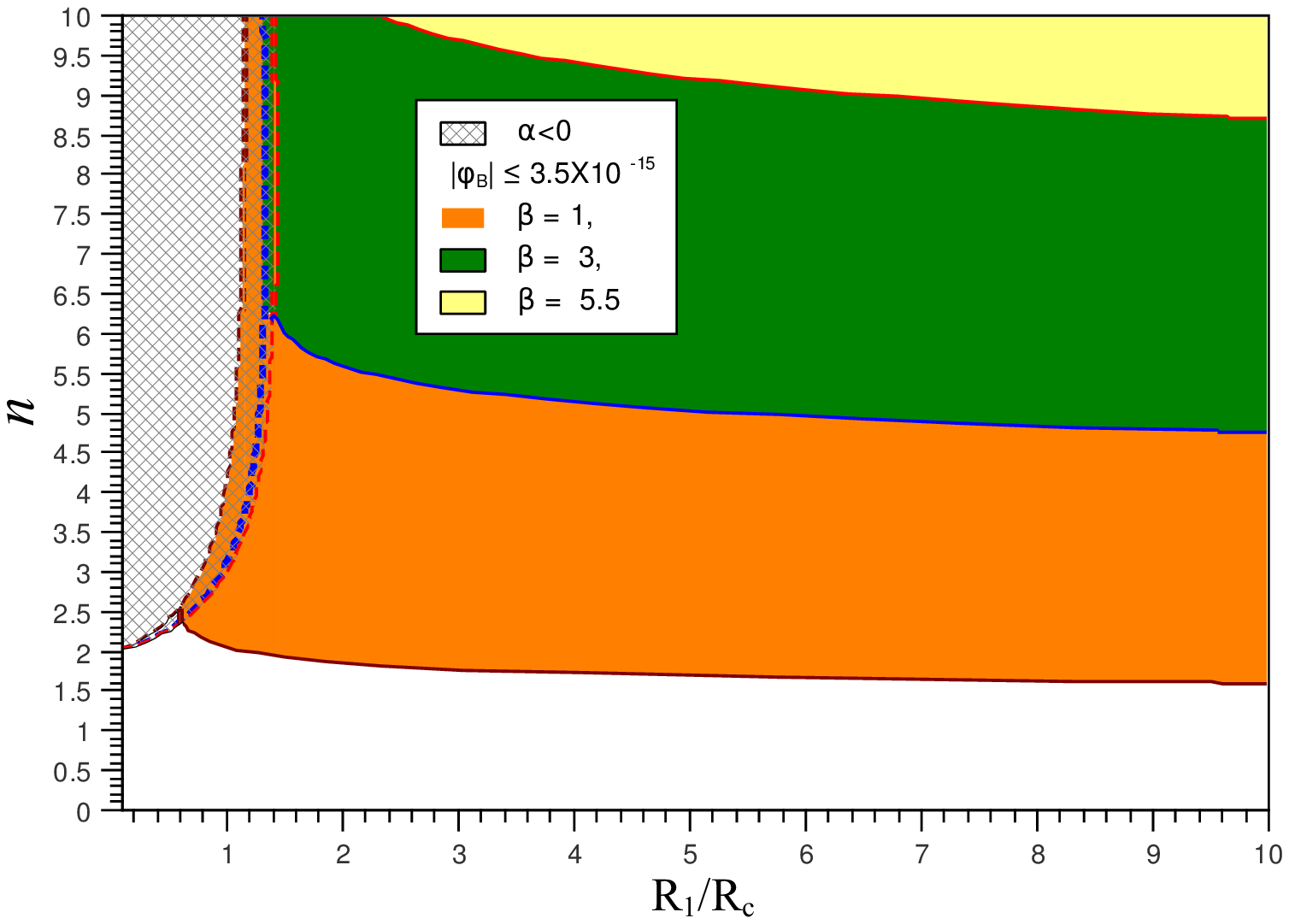}
\caption{The figure shows the allowed regions of the parameter space
which satisfy the thin shell condition corresponding to EP
constraint for the various values of $\beta$
 in case $R_1/R_c$ and $n$
  range from 0.1 to 10 and from 0.05 to 10 respectively.
It is clearly seen that  $n/\beta \gtrsim 1.7$ to satisfy local
gravity constraints.} \label{genb}
\end{figure}
So far we have focussed on large $\beta$ limit of Starobinsky model
as singularity is clearly avoided in this case. A comment on the
finite $\beta$ behavior of the model is in order. In this case, the
analysis requires numerical treatment. In Fig.3, we have displayed
the parameter region consistent with local gravity constraints in
case of finite values of $n$ and $\beta$. In agreement with
Ref.\cite{capozziello}, we find that the local gravity constraints
are satisfied provided that $n/\beta \gtrsim 2$, see Fig.3.
Furthermore, in $f(R)$ gravity, the power spectrum acquires an
additional slope \cite{star} which is constrained in Ref.
\cite{tegmark}. As demonstrated  by Starobinsky , $n/\beta$ should
satisfy the constraint, $\frac{n}{\beta}\gtrsim 4$. Thus the model
proposed in Ref.\cite{waga}, with $\beta \to \infty$, violates this
constraint too. If we adhere to observational constraints imposed by
local gravity constraints, the model is vulnerable to curvature
singularity. It is really interesting that the height of the barrier
between de-Sitter minimum and curvature singularity turns out to be
proportional to $\beta$ which is heavily constrained by local
gravity constraints. In what follows, we shall address this issue.
\section{Finite time generic singularity and its reconciliation}
Let us note that in the limit of $R \to \infty$, $\Delta_{,R} \to 0$
and  the maximum of the potential is located at $\psi = 1$ whose
magnitude is given by
\begin{equation}
\label{maxv}
 V = \frac{R\Delta_{,R} - \Delta}{2(1+\Delta_{,R})^2} \Big |_{R = \infty},
\end{equation}
Since $\lim_{R \to \infty} R\Delta_{,R} = 0$ and $\lim_{R \to \infty}
\Delta = -\alpha \beta R_c$, we find that $\lim_{R \to \infty} V/R_c
= \frac{\alpha \beta}{2}$. The minimum of the potential in free
space  given by (\ref{maxv})
 can be estimated numerically,
${V_{min}}/{R_c} \simeq \mathcal{O}(1)$ at $R\simeq R_1$. In this
case the height of the potential barrier for large value of $\beta$
is approximately  equal to $\beta/2$ as shown in the Fig.\ref{pott}.
The local gravity constraints impose a restriction on the height of
the barrier or equivalently, the parameter $\beta$ for a given value
of $n$ and $R_c$. In case of $n=2$, $R_1/R_c=4$ and $\beta=1$, the
model passes both the local gravity constraints. For large values of
$\beta$, the height of the potential barrier becomes large thereby
hiding the singularity but resulting into clear violation of local
gravity constraints. We also observe that taking small values of
$\beta$, the de-Sitter minimum shifts towards singularity, see
Fig.\ref{pott}. This implies that we should have moderate values of
parameters for a viable evolution. Situation gets worse when we move
to high density regime whose treatment requires extreme fine tuning
of initial conditions of the field\cite{whu}.

As we have seen that the size of $|\phi_{min}| \approx |\Delta_{,R}|_{R =
\rho}$ for any viable $f(R)$ gravity and is constrained to be less
than $\mathcal{O}(10^{-10})$. This means that the  minimum of the
potential corresponding to $\psi = 1 + \Delta_{,R}$, is very close to $
\psi = 1$ even in the case of baryonic/dark matter density.

It should be emphasized that in case of large curvature, the quantum
effects become important leading to higher curvature corrections.
Keeping this in mind, we can incorporate $\mu R^2/R_c$ term in the
model under consideration\cite{star,TH}(see Ref.\cite{Abd} on the
similar theme) which allows us to move the singularity away from
$\psi = 1$. The Big Bang nucleosynthesis constraint at $T \sim$ MeV
($z \sim 10^{10})$ or $R \sim 10^{30} \rho_c$ tells us that the
correction term should satisfy the following condition\cite{zhang},
\begin{equation}
 \frac{\mu}{R_c}R^2 \ll R.
\end{equation}
If we choose $R_c \sim \rho_c$, we find that $\mu \ll
\mathcal{O}(10^{-30})$. In case of neutron star with $\rho \sim
10^{43}\rho_c$, the parameter $\mu$ is constrained to be
$\mu<<10^{-43}$. The  local gravity constraints are satisfied in
this case as
\begin{equation}
 |\phi_B|~~\textrm{(from $\mu R^2/R_c$ term)}~~ \sim ~~ \Delta_{,R}|_{R=10^5R_c}
\sim 2\frac{\mu}{R_c}10^5R_c \ll \mathcal{O}(10^{-38}).
\end{equation}
Let us note that in case we intend to describe inflation with the
help of $R^2$ terms $-$ {\it a la} Starobinsky model, the numerical
value of $\mu$ is much smaller than the quoted value. Indeed, the
mass of scalaron ($R_c^{1/2}/6 \mu^{1/2}$), if it is to be inflaton,
should be $10^{-6} M_p$\cite{star} which implies that $\mu$ is much
smaller than its numerical value quoted in case of neutron star.
Such a correction does not disturb the neutron star physics and
nucleosynthesis constraint but can help in avoiding the curvature
singularity. As a result, the correction term can not contribute any
effect to the local gravity experiments. This implies that the
behavior of the model improves in the high curvature regime though
it might be quite challenging to probe it numerically.

\begin{figure}
\includegraphics[width=11cm]{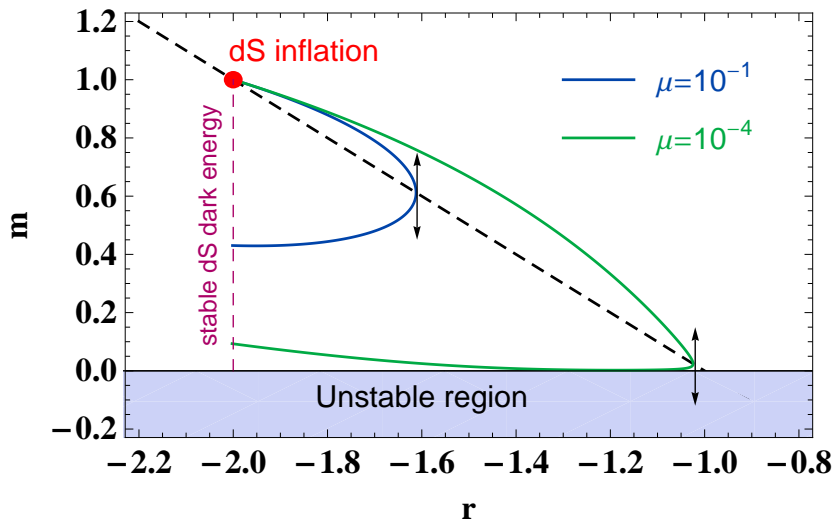}
\caption{The $(r,m)$ plane for the HSS model ($\beta = 1, n = 2$) with the additional term $\frac{\mu}{Rc}R^2$. The dashed diagonal line is the critical line $m=-r-1$. The numerical value of
$\alpha$ is chosen such that the condition of the stability of the future de Sitter stage is satisfied.}
\label{mr}
\end{figure}
We can see from Fig.(\ref{mr}) that in case the BBN condition on
$\mu$ is satisfied, the model would have a standard matter phase.
Then a small $\mu$ is not only necessary for the BBN but also for
the matter phase. Furthermore because of this additional term the
curve $m(r)$ cross the line $m=-1-r$ for a finite $R$ while
$R=\infty$ for the HSS model. Thus we can connect the early phase of
accelerated expansion to the late time acceleration of universe
without a singularity of the curvature scalar $R$.

\section{Conclusions}
In this paper we have examined the variants of HSS models described
by three parameters, $\alpha$, $\beta$ and $n$\cite{waga}. The HSS
scenario in Starobinsky parametrization  corresponds to $n=2$ and
$\beta \lesssim 1$. These models can satisfy the local gravity
constraints and have potential capability of being distinguished
from cosmological constant. The de-Sitter minimum of the effective
potential of the scalar degree of freedom is quite close to
curvature singularity for moderate values of $\alpha$ and $\beta$ in
these models. For larger values of $\alpha$ and $1/\beta$, the
de-Sitter minimum  moves towards singularity. The potential barrier
between the de-Sitter minimum and curvature singularity is finite in
HSS models which makes them delicate. Thus one should carefully tune
the scalaron  evolution such that it does not hit singularity while
evolving in the neighborhood of the minimum of effective potential.
For a given value of $n$, the height of the barrier is defined by
the parameter $\beta$ which is large for larger values of $\beta$
thereby hiding the singularity behind the potential
barrier\cite{waga}. However, large potential barrier between
singularity and de-Sitter minimum comes into conflict with the local
gravity constraints.

The high curvature behavior of $\Delta_R$ is extremely crucial for
local gravity constraints to be evaded. In case $\Delta_{,R}\to 0$
slowly as it happens in case of large $\beta$, we can not satisfy
the local gravity constraints. On the contrary, if $\Delta_R$
approaches zero fast, the corresponding models become more
vulnerable to singularity as the minimum of the effective potential
moves very near  to singularity in the high curvature regime. In
this case, the models under consideration can hardly be
distinguished from cosmological constant.

 The
HSS scenario is build very carefully such that the curvature
dependence of $\Delta(R)$ is just right to satisfy the local gravity
constraints and at the same time to allow  to distinguish itself
from $\Lambda CDM$. Thus the finite time singularity in viable
$f(R)$ models is generic. However, the safe passage of the scalaron
to the minimum of its effective potential can be ensured by the
appropriate fine tuning of the scalaron evolution\cite{whu}. The
fine tuning turns ugly in case of compact objects like neutron
stars. The introduction of higher curvature terms becomes legitimate
near singularity and can in principle improve the behavior of the
model. The resulting scenario can give rise to a viable cosmic
evolution.
\section{Acknowledgements}\emph{}
We are indebted to A. Starobinsky  for taking pain in reading
through the first draft of the manuscript and making important
suggestions for its improvement. We also thank L. Amendolla and I.
Waga for useful discussions. RG thanks the Centre for Theoretical
Physics, Jamia Millia Islamia, New Delhi for hospitality. IT is
supported by ICCR fellowship. MS is supported by ICTP through its
associateship program.

\end{document}